\documentclass[aps,prl,twocolumn,superscriptaddress,showpacs,floatfix]{revtex4-1}
\usepackage{epsfig}
\usepackage{amsmath}

\newcommand{\be}{\begin{equation}}
\newcommand{\ee}{\end{equation}}
\newcommand{\bc}{\begin{center}}
\newcommand{\ec}{\end{center}}
\newcommand{\bi}{\begin{itemize}}
\newcommand{\ei}{\end{itemize}}
\newcommand{\ba}{\begin{eqnarray}}
\newcommand{\ea}{\end{eqnarray}}

\newcommand{\ignore}[1]{}

\begin{document}
\draft
\title{Brain organization into resting state networks emerges at criticality on a model of the human connectome}.

\author{Ariel Haimovici}\affiliation{Departamento de F\'{i}sica, Facultad de Ciencias Exactas y Naturales, Universidad de Buenos Aires, Buenos Aires, Argentina}
\affiliation{Consejo Nacional de Investigaciones Cient\'{i}ficas y Tecnol\'{o}gicas, Buenos Aires, Argentina.}

\author{Enzo Tagliazucchi}\affiliation{Neurology Department and Brain Imaging Center, Goethe University, Frankfurt am Main, Germany.}
 
\author{Pablo Balenzuela}\affiliation{Departamento de F\'{i}sica, Facultad de Ciencias Exactas y Naturales, Universidad de Buenos Aires, Buenos Aires, Argentina}
\affiliation{Consejo Nacional de Investigaciones Cient\'{i}ficas y Tecnol\'{o}gicas, Buenos Aires, Argentina.}

\author{Dante R. Chialvo}
\affiliation{Consejo Nacional de Investigaciones Cient\'{i}ficas y Tecnol\'{o}gicas, Buenos Aires, Argentina.}
\affiliation{Facultad de Ciencias M\'{e}dicas, Universidad Nacional de Rosario, Rosario, Argentina}
\affiliation{David Geffen School of Medicine, University of California Los Angeles, Los Angeles, CA, USA}

\date{\today}

\begin{abstract}

The relation between large-scale brain structure and function is an outstanding open problem in neuroscience. We approach this problem by studying the dynamical regime
 under which realistic spatio-temporal patterns of brain activity emerge from the empirically derived network of human brain neuroanatomical connections. 
The results show that critical dynamics unfolding on the structural connectivity of the human brain allow the recovery of many key experimental
findings obtained with functional Magnetic Resonance Imaging (fMRI), such as divergence of the correlation length, anomalous 
scaling of correlation fluctuations, and the emergence of large-scale resting state networks.
\end{abstract}
\maketitle

Understanding the relation between brain architecture  and function is a central question in neuroscience.  In that direction, important efforts over recent years have been devoted to map the large-scale structure of the human cortex, including  attempts to build brain structural connectivity matrices from imaging data.  An example is the connectivity matrix of the entire human brain, recently derived from fiber densities measured between a large number (500-4000) of homogeneously distributed brain regions \cite{Hagmann08}. This and related work encompasses a large collaborative project dubbed the brain ``connectome" \cite{conectoma}, whose ultimate goal is to understand in detail the architecture of whole-brain connectivity. 
However, ``like genes, structural connections alone are powerless'', thus ``the connectome must be expressed in dynamic neural activity to be effective in behavior and cognition'' \cite{Spornsbook}.  
The results presented in this Letter show that very relevant aspects of brain dynamics can be predicted from the structure provided that the underlying dynamics are critical.

To guide our comparison with available experimental results, we choose to concentrate on robust findings concerning brain dynamics. Specifically, 
we ask how spontaneous brain dynamics at the large scale organize into the relatively few spatio-temporal patterns revealed experimentally in 
recent years \cite{RSN}. This is important because a wide range of experiments using functional Magnetic Resonance Imaging (fMRI) have 
emphasized that these spatial clusters of coherent activity, termed Resting State Networks (RSN) \cite{beckmann05}, 
are specifically associated with  neuronal systems responsible for sensory, cognitive and behavioural functions \cite{Smith}. 
Furthermore, the pattern of correlations in these networks has been shown to change with various cognitive and pathophysiological conditions \cite{RSN}.
Of interest here are studies showing that the RSN activity exhibits peculiar scaling properties, resembling dynamics near the critical point of a second order phase transition \cite{bak,chialvo10,Expert11}, consistent with evidence showing that the brain at rest is near a critical point \cite{Tagliazucchi12}.  These empirical findings are in line with computational modeling results \cite{Fraiman09,Kitzbichler09, Deco12}.

\begin{figure}[htb]
\begin{center}
\includegraphics*[width=0.7\columnwidth,clip=true]{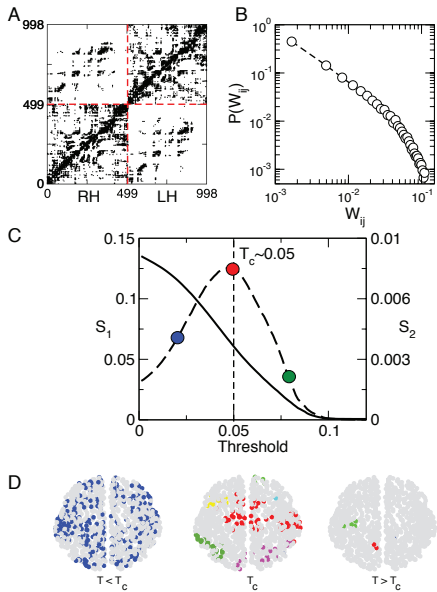}
\end{center}
\caption{\label{fig1}The model. Panel A shows the model adjacency matrix (i.e., the connectome) and Panel B the edge weights distribution. Data obtained from the white matter tracts connecting a parcellation of 998 nodes covering the entire human brain \cite{Hagmann08}. RH and LH refers to the right and left brain hemisphere, respectively.  Panel C shows the giant cluster's size, (i.e., the order parameter, $S_1$, solid line) and the second largest cluster's size ($S_2$, dashed line) as a function of the threshold $T$ (control parameter) as well as the critical point $T_c \sim 0.05$. Panel D illustrates examples of clusters (denoted with different colors) at the three $T$ values denoted with colored markers in Panel C.}
\end{figure}
 
Here we study whether a simple dynamical model running over the empirical structure of neuroanatomical connections \cite{Hagmann08} suffices to replicate the aforementioned fundamental features of spontaneous brain activity repeatedly seen in fMRI experiments. The model consists of a network of interconnected nodes (i.e., the connectome) , together with a dynamical rule. The matrix of connections follows the neuroanatomical connectivity described recently by Hagmann et al. \cite{Hagmann08},  who studied  healthy human subjects and reported the average fiber tract density between any two brain areas (from a gray matter parcellation into 998 areas).
To complete the model we need to specify the dynamics of each node. For simplicity, we choose  discrete state excitable dynamics following the Greenberg-Hastings model \cite{Greenberg}. Thus, each node  is assigned one of three states: quiescent $Q$, excited $E$, or refractory $R$,  and the transition rules are: 1) $Q \rightarrow E$  with a small probability $r_1$ ($ \sim 10 ^{-3}$), or if the sum of the connection weights $w_{ij}$ with the active neighbors ($j$) is higher than a threshold $T$, i.e., $\sum w_{ij} > T$ and $Q  \rightarrow Q$  otherwise; 2) $E \rightarrow R$  always; 3) $R \rightarrow Q$ with a small probability $r_2$ ($\sim 10 ^{-1}$)  delaying the transition from the $R$ to the $Q$ state for some time steps. We held fixed parameters $r_1$ and $r_2$, which determine the time scales of self-excitation and of recovery from the excited state, respectively, and changed $T$.
For the numerical analyses, the time series of each node was binarized  by assigning state $E=1$ and the remaining  states into 0's and convolved  with a standard hemodynamic response function \cite{hrf} mimicking the  brain neuro-metabolic coupling.

As depicted in Figure 1, the dynamics of the model show a transition as a function of the threshold $T$. For relatively small values of $T$ even the weakest connections are enough for the activity to spread, resulting in a regime with a relatively high activity level. On the contrary, for high values of $T$ the activity only flows through the few strongest connections and therefore the overall  activity decreases. 
To characterize the transition between these regimes an order parameter was defined considering the sizes of the active clusters. Clusters are groups of nodes simultaneously activated and linked to each other through 
a non zero $w$. At each time step the size of the largest cluster ($S_1$) and the second largest cluster ($S_2$) were computed. These calculations (see Panel C of Fig. \ref{fig1}) unveil a transition between a phase in which a giant cluster covers $\sim 15\% $ of the system (while the second largest cluster is of negligible size) and another phase in which only scarce  activations occur and  the nodes fail to coalesce into large clusters. At an intermediate value, a critical point ($T_c$ in Fig. 1C) can be identified by the peak in the size of the second largest cluster, as done usually in percolation \cite{Stauffer} as well as recently in human fMRI experiments \cite{Tagliazucchi12}. The finite size of the available connectome makes the usual demonstration of criticality in the thermodynamical limit impractical,  thus a range of alternative indicators are provided instead (see also Supp. Info.).

We compare now the dynamics of the model with previous experimental results, in particular, with two robust features exhibited by the spontaneous activity of human brain RSN \cite{Fraiman12}: 1)  the correlation length of brain activity increases with size (as expected by the divergence of correlation length), and 2) the variance of the short-term correlations  between pairs of brain sites remains high,  independently of the number of pairs considered. Since these two properties are often seen as generic features of criticality, we decided to explore first whether the model exhibits similar dynamics.
To compare with the experimental results, each node of the model was labeled as belonging to the closest human RSN by matching the node's coordinates provided by Hagmann et al \cite{Hagmann08} with a spatial mask of the RSN \cite{beckmann05,Fraiman12}. Nodes that were farther than 1 cm from the closest RSN were discarded from the analysis (see Supp. Info.).
\begin{figure}[htb]
\begin{center} 
\includegraphics*[width=0.85\columnwidth,clip=true]{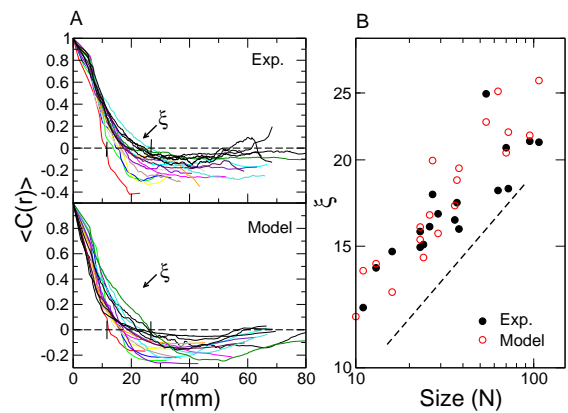}
\end{center}
 \caption{The correlation length $\xi$ of the activity in the model near $T_c$ increases with the cluster size (N), as reported for human brain data \cite{Fraiman12}. Panel A shows the correlation function $C(r)$ computed from human data (Exp.) and from the model at $T_c$ (colored lines are used for the different clusters).  The correlation length $\xi$ is the distance $r$ where $C(r) = 0$, (range denoted with the arrows). Panel B shows the $\xi$ values for the functions plotted on panel A, demonstrating that $\xi \sim N^{1/3}$ (dashed line),  both in the experiment and model data.}
\end{figure}

\emph{Divergence of the correlation length.} The correlation length represents the average distance at which two points in the system behave 
independently, and is known to diverge at criticality \cite{Cavagna}. Following a standard procedure  \cite{Fraiman12,Cavagna}, we computed the average correlation function of the signal fluctuations between all pairs of nodes in each cluster which are separated by a distance $r$,  yielding the correlation function $C(r)$ (see Supp. Info.).

Fig. 2A corresponds to the two-point correlation function as a function of distance for all cluster sizes obtained experimentally and in the model at $T_c$ . Panel B shows the dependence of $\xi$ with the cluster size $N$.    Both numerical estimations clearly show that near $T_c$  the divergence of the correlation length found in the experiments \cite{Fraiman12} is reproduced by the model.

 \begin{figure}[htb]
 \begin{center} 
\includegraphics*[width=0.92\columnwidth,angle=-90,clip=true]{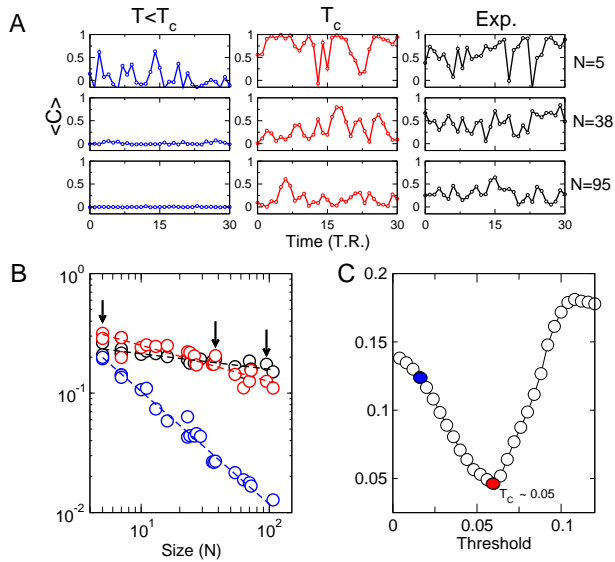}
\end{center} \caption{The short-term correlation $<C>$ in the model at $T_c$ exhibits transient fluctuations at all cluster sizes as seen in human brain data \cite{Fraiman12}. Panel A: Examples of $<C>$  fluctuations for three cluster sizes $N$ at critical ($T_c$) and subcritical ($T<T_c$) dynamical regimes of the model, as well as for the human brain data (Exp.). Panel B shows that the variance of the fluctuations in  $<C>$ remains approximately constant only for the human brain data (empty black circles) and for the model at $T_c$ (filled red circles). For $T< T_c$ (filled blue circles) the variance decreases faster with N. The three small arrows denote the sizes used in the examples of Panel A.  Panel C shows a plot of the distance between the scaling of the fluctuations of the human fMRI data and those from the model for a wide range of $T$. Notice that the best agreement occurs for $T_c$. Colored markers in Panel C correspond to the $T_c$ and $T< T_c$. values used in Panels A and B.}
 \end{figure}

\emph{Temporal fluctuations of the mean correlation in the RSN.}  As recently shown \cite{Fraiman12}, the time evolution of the correlation within these patterns exhibits bursts of high correlation intermixed with instances of dis-coordination. 
Panel A in Fig. 3 shows examples of the fluctuations in the short-term mean correlation $<C>$ between all pairs of nodes within a given cluster, in this case calculated in non-overlapping time windows of 10 steps. At the critical state, the variance of $<C>$ is of the same order as observed in the experiments for different cluster sizes. 
Panel B shows the dependence of the fluctuations in $<C>$ with the cluster size $N$. At the subcritical regime the fluctuations decrease as $\frac{1}{N}$, which reveals the asynchrony of the active nodes. On the other hand, at the critical state they remain approximately constant,  similar to what is observed in experimental fMRI data \cite{Fraiman12}. In the  supercritical regime, the  vanishingly low level of  activity prevents high correlations and therefore the fluctuation amplitudes are close to zero for all cluster sizes. To compare with the experimental human fMRI data, we computed the root mean square distance between the model (m) and the experimental  data (e) as $\Delta=\sqrt{ \frac{\sum_{N_c} (\sigma_{<C>}^{\text{(e)}} - \sigma_{<C>}^{\text{(m)}})^2 }{N_c} }$ where the sum is over all clusters $N_c$. Panel C shows that the distance is minimum precisely at the critical point of the model.

\begin{figure}[tb]
\begin{center} 
\includegraphics*[width=0.9\columnwidth,clip=true]{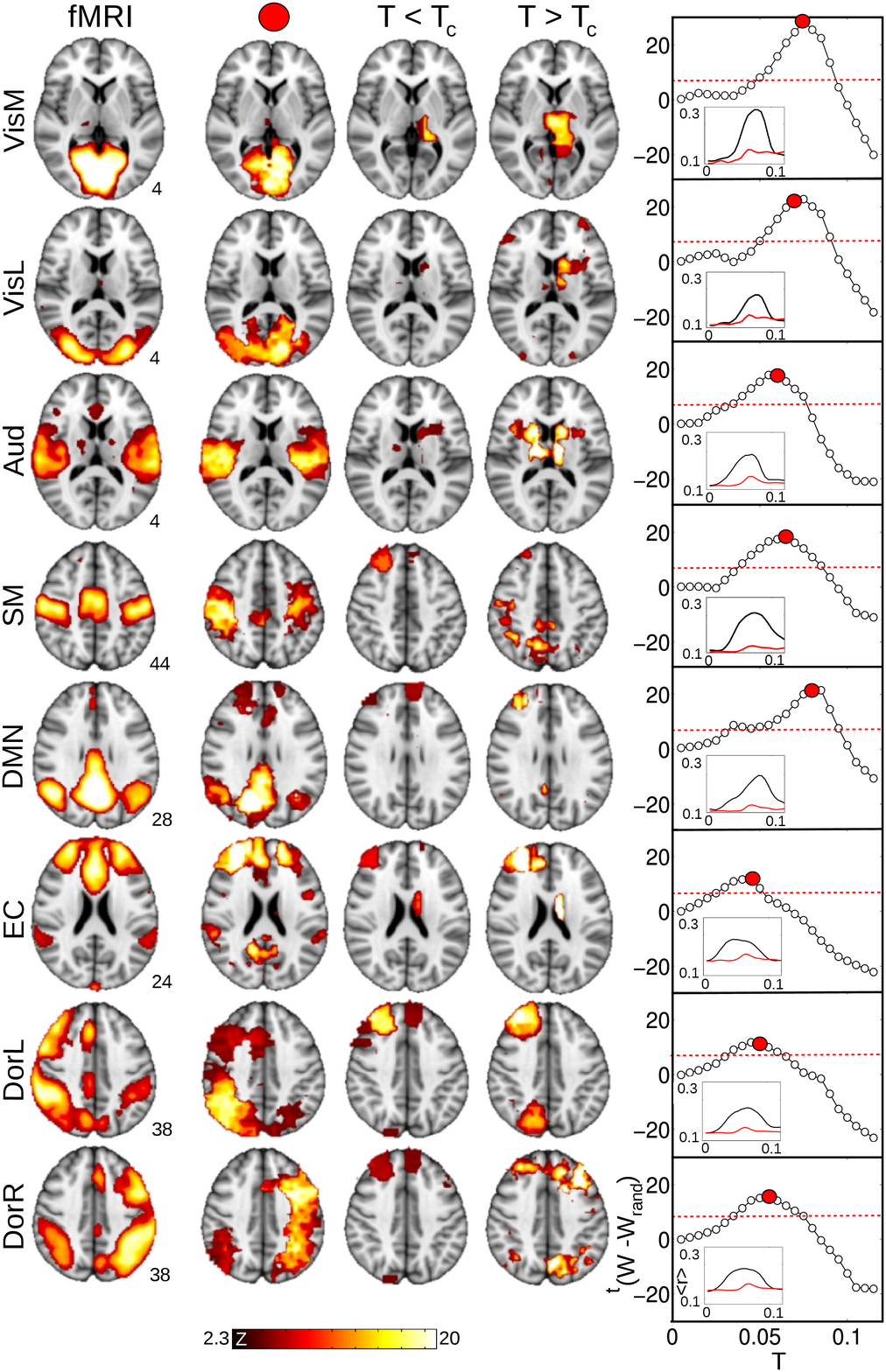}
\end{center}
\caption{\label{fig4}The spatial organization of the  human brain RSN emerges spontaneously in the model near $T_c$. RSN obtained from the model 
with ICA (averaged across all trials) are shown for three values of $T$:  $T <  T_c$ (0.03) , $T > T_c$ (0.1)  and the correlation maxima with the human RSN from \cite{beckmann05} 
(``fMRI"), which is always $\sim T_c$ (red marker). Right column, $t(W-W_{rand})$, shows the T-statistic  for the difference in the correlations between the model using the connectome and a randomized version. The red horizontal line indicates the
level of $p<0.001$, corrected. In the insets, the mean spatial correlation $<r>$ is shown for the real (black) and randomized (red) connectomes
as a function of $T$.
Medial visual (VisM), lateral visual (VisL), auditory (Aud), 
sensory-motor (SM), default mode network (DMN), executive control (EC), dorsal visual  stream left (DorL) and right (DorR). 
Numbers beneath each brain slice denote its horizontal coordinate.}
\end{figure}

\emph{RSN spatial patterns emerge only at criticality.} One of the most revealing features of large-scale spontaneous brain activity is its spatial 
organization into RSN. Their functional relevance is highlighted by the fact that the spontaneous  activity closely parallels brain activation 
patterns seen during task execution \cite{Smith}.
We studied whether these patterns can be seen in the model using the same methods employed to reveal RSN in experimental data (Independent Component
 Analysis -ICA) as implemented in the FSL MELODIC software \cite{beckmann04}. First, we identified the cortical locations of the model's 998 nodes, by parcellating the brain gray matter in cortical patches  via a random growth algorithm (see Supp. Info). Then the  model time series of each region of interest was assigned to the corresponding parcellation patch (plus Gaussian noise of 0.15 times the variance of the signal), and an ICA decomposition into 8 independent components was done (100 trials for each $T$ value). 
For each $T$, we computed the maximal spatial correlation between the location of the model ICA components and the spatial maps of a set of well-established human RSN \cite{beckmann05}.  The statistical significance of these findings was explored by computing T-statistics against a null hypothesis constructed with a randomized version of the connectome. In all cases, the model dynamics near $T_c$  best replicates the empirical findings (see Fig. 4). This implies that the RSN coordinated spontaneous activity unfolds at the same anatomical locations both in the human brain and in the model close to $T_c$ ,  something already evident by visual inspection of the patterns presented in Fig. 4.

\emph{Discussion.} This is the first demonstration that a hybrid modeling approach (realistic anatomical connectivity plus a simple dynamical rule) suffices to capture relevant spatio-temporal aspects of brain dynamics, provided that the dynamical regime is critical. These aspects include generic features of critical systems, but also the emergence of structures having a well-established neurobiological meaning, namely the cortical RSN. While experimental evidence for the aforementioned signatures of criticality in brain systems was already discussed \cite{chialvo10}, here we made a stronger point: in the model, the critical regime 
appears as a necessary condition for the emergence of neurobiologically relevant aspects of brain dynamics.
Our result also represents an important first step in the direction of realistic hybrid computational modeling of large-scale brain function both in health and disease. As an example,
many altered brain states are associated with RSN alterations, a prominent example being the loss of consciousness in the comatose state \cite{norton}. 
In light of the present results such brain state alterations can be regarded as a displacement from an optimal dynamical point.

Summarizing, the results show that by endowing with critical dynamics the brain network of anatomical connections (or connectome),  key observations
 about brain dynamics can be replicated. These results contribute
 to close the gap between structural and functional network connectivity in the human brain, by emphasizing the dynamical regime at which models should 
predict a wide range of observations about large scale brain function.
 
Supported by CONICET (Argentina), by  NIH (USA)  and by
LOEWE NeFF and BMBF (Germany). We thanks O. Sporns (Indiana Univ.) for sharing the connectome data and C. Beckmann (Imperial College) for the masks reported in \cite{beckmann05}.

\end{document}